# Ordered InAs quantum dots on pre-patterned GaAs (0 0 1) by local oxidation nanolithography

J. Martín-Sánchez, Y. González, L. González, M. Tello, R. García, D. Granados, J.M. García, F. Briones

*Instituto de Microelectrónica de Madrid (CNM-CSIC), MBE, C/Isaac Newton, 8 (PTM), 28760 Tres Cantos, Madrid, Spain*

Abstract

Ordered InAs quantum dot (QD) arrays have been obtained on pre-patterned GaAs (0 0 1) substrates by atomic force microscopy (AFM) local oxidation nanolithography. Prior to InAs molecular beam epitaxy (MBE) deposition, an ordered square array of nanoholes is formed at the GaAs pre-patterned surface following in situ etching with atomic hydrogen. A low substrate temperature is maintained during the whole process in order to avoid pattern smoothing. Our results show that the density and dimensions of the nanoholes on the GaAs surface determine InAs QD size, nucleation site and InAs dose necessary for their formation. As a function of the geometrical parameters of the nanohole array, we can obtain either ordered 2D arrays of separated QD, closely packed QD or localized areas for QD formation.

1. Introduction

In spite of large efforts, no reliable procedure has been described up to now to convince self-assembled InAs/GaAs (0 0 1) quantum dots (QD) to self-organize also into a regular array with homogeneous size and distance distribution as it is the case for other epitaxial systems, such as PbSe/Pb$_{1-x}$Eu$_x$Te [1].

However, as advanced optoelectronic devices and new applications in nanoelectronics, like logic devices and memories, would clearly benefit from such ordered and precisely situated nanostructures, different approaches have been tried to artificially force QD to self-assemble at predefined sites on pre-patterned substrates [2], [3], [4], [5], [6], [7], [8], [9], [10]. In this situation, it is vital that the lithographic process used to pre-pattern the substrate will not introduce dislocations, impurities or other point defects at the QD nucleation site in order not to degrade the QD electronic and optical properties. In this sense, the use of atomic force microscopy (AFM) local oxidation has been demonstrated recently as a powerful patterned fabrication technology to obtain ordered arrays of InAs/GaAs and InAs/InP QD [4], [5], [10]. Following this approach, in this work we present a process that results in ordered InAs QD arrays on GaAs (0 0 1) substrates pre-patterned with periodic arrays of nanometre size oxide dots produced by AFM local oxidation [11], [12], [13], [14], [15]. In this work we study the InAs QD nucleation on patterns with very different geometrical parameters, from completely isolated to fully collapsed features. In this way we try to control the precise formation site of each nanostructure, which is of clear interest in the development of single photon emitters [16], and to obtain ordered 2D arrays of nanostructures with high density for its application in QD lasers [17]. Our results show that the density and dimensions of the nanoholes on the GaAs surface determine InAs QD size and its nucleation site. As a function of the geometrical parameters of the nanohole array, we can obtain either ordered 2D arrays of separated QD, closely packed QD or localized areas for QD formation.

2. Experimental procedure

In our experiments, Si-doped epi-ready GaAs (0 0 1) substrates were patterned by local oxidation nanolithography using a non-contact AFM in a controlled humidity environment [11]. Different 2D arrays (labeled A, B and C) of GaAs oxide dots were fabricated on the same substrate. The particular geometrical parameters of these patterns, GaAs oxide dots diameter (φ), pitch distance (λ), φ/λ ratio and oxide dots density (ρox) are shown in detail in Table 1. Notice that a value of φ/λ≈0.6 implies well-separated GaAs oxide dots in the pattern, φ/λ=1 implies that the GaAs oxide dots are just touching each other, and for φ/λ>1 patterns, the GaAs oxide dots are overlapping. φ and λ parameters are schematically represented in Fig. 1.

| GaAs oxide dot pattern | φ (nm) | λ (nm) | φ/λ | ρox (cm$^{-2}$) |
| --- | --- | --- | --- | --- |
| A | 95 | 165 | 0.6 | $3.7 \times 10^9$ |
| B | 30 | 30 | 1 | $110 \times 10^9$ |
| C | >λ | 10 | >1 | $650 \times 10^9$ |

Table 1. Geometrical parameters of different 2D arrays of GaAs oxide dots (labelled A, B and C) fabricated on the same substrate: GaAs oxide dot diameter (φ), pitch distance (λ), φ/λ ratio and density (ρox) of oxide dots.

The aim of this work is to use the nanoholes, left on the GaAs patterned surface after oxide desorption and buffer growth, as nucleation centres for InAs QD. The main restriction is the maximum temperature used in the whole process, which should be as low as possible in order to avoid smoothing of the pattern profile [6]. Our approach is to remove oxides at low temperature using atomic hydrogen and to grow a thin buffer layer also at low temperature by using atomic layer molecular beam epitaxy (ALMBE) growth technique [18].

Both GaAs patterned oxide and surface native oxide were etched away in the molecular beam epitaxy (MBE) chamber by exposing the surface to an atomic hydrogen flux using a Ta H$_2$ thermal cracker with a H$_2$ base pressure of $10^{-5}$ mbar during 5 min at substrate temperature ($T_s$) of 450 °C. After oxide removal, a thin GaAs buffer layer (12 monolayers, ML) was grown at 1 ML/s by ALMBE [18] also at Ts=450°C. Then, InAs was deposited at Ts=490°C using a standard growth sequence of 0.1 ML of InAs deposition at 0.07 ML/s followed by a pause of 2 s under As$_2$ flux.

3. Results

First, we have studied in detail the oxide removal process by exposure GaAs (0 0 1) flat unpatterned surfaces to atomic H flux. After 3 min under H flux at Ts=450°C, we observe in situ, by reflection high-energy electron diffraction (RHEED), a (1×1) diffraction pattern and after 5 min H exposure, a clear (2×4) pattern is developed. Subsequent ex situ AFM measurements of surface roughness results in a rms value of 0.3 nm showing that atomic hydrogen exposure provides much smoother surfaces than those produced by conventional oxide thermal desorption (5 min at Ts=630°C under As$_4$ flux) where AFM rms roughness values of 1.8 nm are typically obtained in our laboratory.

After oxide removal, growth at Ts=450°C of 12 ML thick GaAs buffer layer by ALMBE proceeds without any observable RHEED pattern degradation. When InAs is deposited on the sample at Ts=490°C, we observe a thickness for 2D–3D transition ($h_{2D-3D}$) of 1.7 ML, as conventionally obtained for self-assembled QD formation under our experimental conditions.

Fig. 1a shows an AFM image and profile of part of a (10×10) 2D array of oxide dots obtained by non-contact AFM local oxidation. Fig. 1b shows an AFM image and profile of the nanoholes obtained in the same patterned area presented in Fig. 1a after oxide removal and subsequent 12 ML GaAs buffer layer growth at Ts=450°C. It is clearly seen that the above low-temperature process maintains the periodic array of nanoholes. We also observed that the nanoholes show distinct concentric features: a deep central hole surrounded by a cavity with smoother slope walls, in correspondence with the previous oxide thickness distribution. From this point on, we will talk of outer and inner holes to distinguish these features.

The next step is InAs deposition for QD formation. Assuming that InAs preferentially nucleates in the nanoholes, the amount of InAs deposited should be adjusted to the density of the pattern in order to avoid InAs dots formation outside the patterned areas. In our experiments we have limited the total amount of deposited InAs to only 0.5 ML, far from critical thickness (1.7 ML) for self-assembled QD formation on GaAs (0 0 1) flat surfaces. RHEED diagram remains 2D after 0.5 ML InAs deposition, because AFM patterned area is much smaller than the RHEED electron beam illuminated area and therefore, post-growth AFM characterization is the only available tool in order to monitor QD formation on patterned areas at subcritical thickness.

For 2D arrays with $\varphi/\lambda \approx 0.6$, InAs QD appear inside the GaAs nanoholes as it is clearly observed in Fig. 2a. This demonstrates that GaAs nanoholes effectively act as preferential nucleation sites for the deposited InAs. Obviously, Stranski–Krastanov mechanism is not responsible for QD formation. Moreover, InAs nucleation takes place at the inner hole, as it is clearly observed in the 3D zoom shown in the inset of Fig. 2a. One, two or three InAs QD within the same nanohole can be observed. QD molecules have been reported by other authors in patterned substrates obtained by other techniques [19], [20]. In our case, we find that the number and geometrical display of the QD can be correlated with the particular shape of the oxide dots previous to nanohole formation. In fact, we have observed that the GaAs oxide dots have sometimes one, two or three lobes. As shown in Fig. 2b, the number and orientation of the InAs QD formed after the growth process (right column of Fig. 2b) replicate the number and orientation of the oxide dot lobes (left column of Fig. 2b). Thus, in order to obtain reproducible 2D arrays of InAs QD it is mandatory a subnanometre control of the local GaAs oxidation process.

Another feature observable in Fig. 2a is the formation of some InAs 3D nuclei situated on the free spaces between nanoholes. This can be due to the presence of residual pits or defects at the surface that could also actuate as undesirable nucleation sites for InAs. In order to get rid of these defects, it is necessary to increase the buffer layer thickness or to make the local oxidation patterning on epitaxial layers instead of commercial epi-ready GaAs substrates. We are working in this sense at present.

For dense 2D arrays of nanoholes, with $\varphi/\lambda=1$, highly ordered and densely packed InAs features are obtained, as can be seen in the AFM micrograph shown in Fig. 3. Now, various types of InAs features can be identified: first, very small dots (base and height) appear at the bottom of the deep (inner) holes (see zoom in the inset of Fig. 3). These tiny InAs dots cannot grow larger probably due to the build up of a large amount of accumulated compressive strain due to the high density of InAs QD embedded into the GaAs matrix. Second, a regular array of InAs dots touching each other along rows and columns directions of the pattern. These dots grow on the mesas formed by collapse of the outer holes during the oxide removal process. Notice that these dots are formed between GaAs inner holes (see AFM profile from diagonal white line, on the left side of Fig. 3). Third, large QD (standard size) are formed at random sites probably due to the excess InAs not able to incorporate in the matrix (deep holes and mesa areas), even for a small total dose of InAs of 0.5 ML.

Other authors [5] have studied the dependence of the InAs nucleation process on the geometry of the pattern by reducing the distance between pattern features in 1D arrays of oxide dots obtained by AFM local oxidation. They do not observe this dual nucleation behaviour (both in deep holes and mesa areas) when InAs is deposited, and only a coalescence process is observed when the distance between the pattern nanoholes is reduced. Our results show that the 2D display of the pattern features can greatly modify the growth process comparing with its behaviour on 1D isolated patterned motives. This can be due to the interaction between the InAs nuclei through the GaAs matrix in high-density 2D patterns. In fact, an increase of the area density of the pattern leads to an increase of the elastic energy density for the same amount of InAs incorporated at the pattern.

Experimental results obtained for φ/λ>1 (Fig. 4) show that a complete overlapping between nanoholes after oxide desorption takes place. On this overlapped nanoholes area, a periodic array of small dots with λ=10nm (see zoom inset in Fig. 4) with the same period as its corresponding 2D oxide pattern (pattern C) are formed. On top of these small dots random InAs QD appear with a density as low as $2.5 \times 10^9$ cm$^{-2}$, although the InAs deposited is much lower than $h_{2D-3D}$. Once again, the concept of critical thickness for QD formation is not applicable anymore. In this case, the pattern density seems to be too high to permit the formation of two lattices of InAs dots (on deep holes and on mesa areas) as in the φ/λ=1 pattern (see Fig. 3). Unfortunately, AFM resolution is not enough to distinguish if the small dots are formed on deep holes or mesa areas for this pattern with very small lattice unit (pattern C).

Therefore, our results shown in Fig. 2, Fig. 4 suggest that depending on the diameter φ and distance λ between the nanoholes of the 2D arrays fabricated on GaAs substrates, we can control the QD localization at specific sites (φ/λ≈0.6 pattern), achieve a closely packed 2D array QD (φ/λ=1 pattern) or confine their formation to predetermined areas of the surface (φ/λ>1 pattern).

4. Conclusions

We have developed a process that results in ordered InAs QD arrays on pre-patterned GaAs (0 0 1) substrates by AFM local oxidation. Prior to InAs growth, the oxide pattern is removed by interaction with atomic hydrogen followed by growth of a thin buffer epitaxial GaAs layer, in a low substrate temperature process in order to avoid pattern smoothing. Our results show that the density and dimensions of nanoholes on the GaAs surface determine the InAs QD assembling process, controlling its size, nucleation site and the InAs dose necessary for its formation. In this way by changing the geometrical parameters of the nanoholes in the pattern of the GaAs oxide dots, we can obtain either ordered 2D arrays of separated QD, closely packed QD or selective areas in which QD are formed. These results show a promising way to produce engineered density of uniform size QD arrays and to control QD site formation.

Acknowledgements

This work was financed by Spanish MCyT under NANOSELF project (TIC2002-04096) and MAT2003-02655 and by the SANDiE Network of excellence (Contract no NMP4-CT-2004-500101 group TEP-0120).

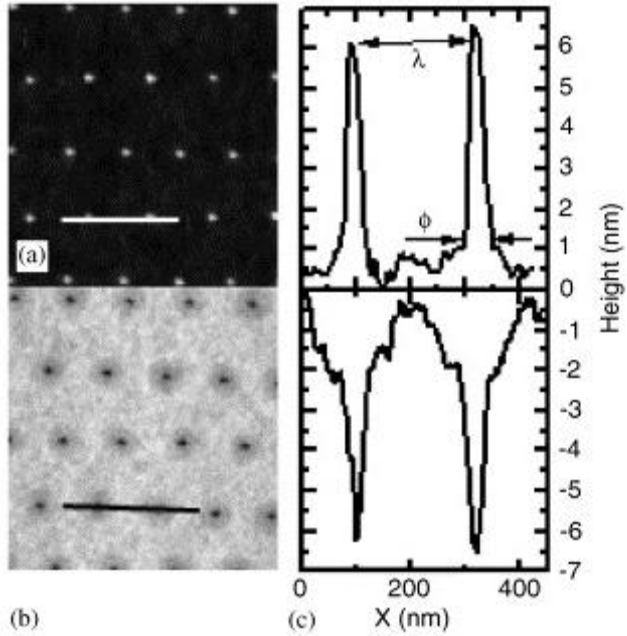

Fig. 1. AFM images of part of a 10×10 2D array of GaAs oxide dots obtained by AFM local oxidation (a) and their corresponding GaAs nanoholes (b) obtained after the GaAs oxides removal and subsequent 12 ML GaAs buffer layer growth. Profiles (obtained along the lines drawn) of GaAs oxides (a) and nanoholes (b) are also shown on the right. GaAs oxide dot diameter (ϕ) and pitch distance (λ) are schematically represented.

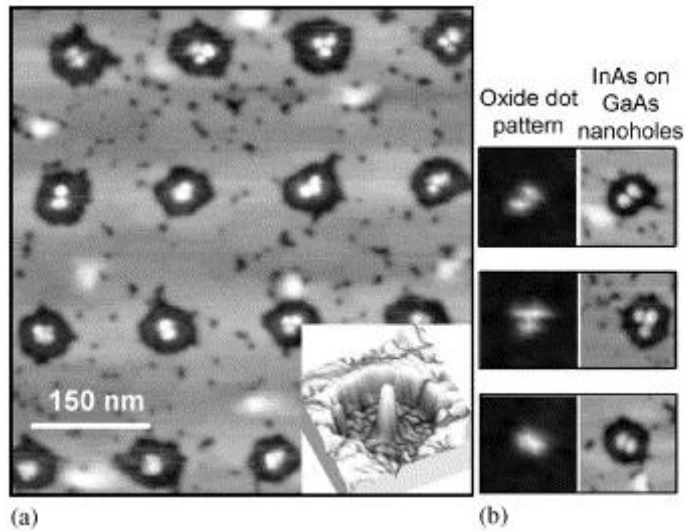

Fig. 2. (a) AFM image of part of the 2D array of QD obtained on GaAs patterned substrates with $\phi/\lambda \approx 0.6$ (ϕ oxide dot diameter and λ pitch distance) after depositing 0.5 ML of InAs. A 3D zoom of one InAs QD inside a GaAs nanohole of the pattern is shown in the inset. (b) Couples of AFM images evidence that the number and orientation of the InAs QD formed after the growth process inside the nanoholes (right) replicate the number and orientation of the oxide lobes (left) in the GaAs pattern.

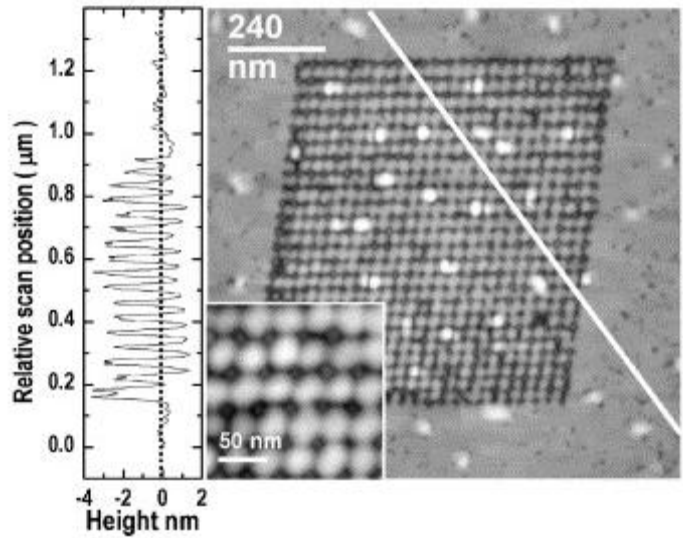

Fig. 3. AFM image of a densely packed InAs QD array obtained by deposition of 0.5 ML of InAs on GaAs patterned substrate with a ratio φ/λ=1 (φ oxide dot diameter and λ pitch distance). A detail of this array can be seen in the zoom shown in the inset at the bottom part. On the left-hand side the profile obtained along the diagonal white line shown in the AFM image is presented.

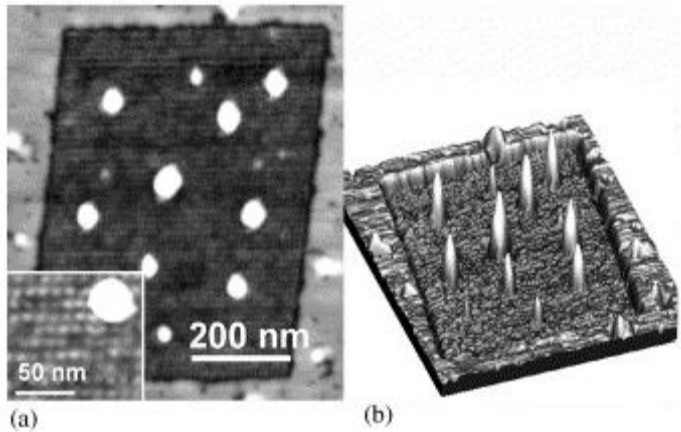

Fig. 4. Two-dimensional and three-dimensional AFM images of InAs QD obtained by deposition of 0.5 ML of InAs on GaAs patterned substrate with a ratio φ/λ>1 (φ oxide dot diameter and λ pitch distance). Notice that the rectangular area defined by the GaAs pattern after oxide desorption acts as preferential nucleation zone for InAs QD.